\documentclass[nofootinbib,superscriptaddress,longbibliography,a4paper,twocolumn]{revtex4-1}
\usepackage{graphicx}
\usepackage{dcolumn}
\usepackage{bm,bbm}
\usepackage{xcolor}
\usepackage{tcolorbox}
\usepackage{algorithm}
\usepackage{algpseudocode}
\usepackage{amsmath}
\usepackage{bbold}
\usepackage{braket}
\usepackage{nameref}
\usepackage[toc,page]{appendix}

\usepackage[colorlinks,breaklinks,linkcolor={blue},citecolor={magenta},urlcolor={blue}]{hyperref}

\makeatletter
\newcommand\org@hypertarget{}
\let\org@hypertarget\hypertarget
\renewcommand\hypertarget[2]{%
  \Hy@raisedlink{\org@hypertarget{#1}{}}#2%
  }
\makeatother

\begin{document} 

\title{Universal quantum computation via quantum controlled classical operations}

\author{Sebastian Horvat}
\email{sebastian.horvat@univie.ac.at}
\affiliation{University of Vienna, Faculty of Physics, Vienna Center for Quantum Science and Technology, Boltzmanngasse 5, 1090 Vienna, Austria}

\author{Xiaoqin Gao}
\email{xgao5@uottawa.ca}
\affiliation{University of Vienna, Faculty of Physics, Vienna Center for Quantum Science and Technology, Boltzmanngasse 5, 1090 Vienna, Austria}
\affiliation{Institute for Quantum Optics and Quantum Information (IQOQI), Austrian Academy of Sciences, Boltzmanngasse 3, 1090 Vienna, Austria.}
\affiliation{Department of physics, University of Ottawa, Advanced Research Complex, 25 Templeton Street, K1N 6N5, Ottawa, ON, Canada}

\author{Borivoje Daki\'c}
\email{borivoje.dakic@univie.ac.at}
\affiliation{University of Vienna, Faculty of Physics, Vienna Center for Quantum Science and Technology, Boltzmanngasse 5, 1090 Vienna, Austria}
\affiliation{Institute for Quantum Optics and Quantum Information (IQOQI), Austrian Academy of Sciences, Boltzmanngasse 3, 1090 Vienna, Austria.}

\begin{abstract}
A universal set of gates for (classical or quantum) computation is a set of gates that can be used to approximate any other operation. It is well known that a universal set for classical computation augmented with the Hadamard gate results in universal quantum computing. Motivated by the latter, we pose the following question: can one perform universal quantum computation by supplementing a set of classical gates with a quantum control, and a set of quantum gates operating solely on the latter? In this work we provide an affirmative answer to this question by considering a computational model that consists of $2n$ target bits together with a set of classical gates controlled by log$(2n+1)$ ancillary qubits. We show that this model is equivalent to a quantum computer operating on $n$ qubits. Furthermore, we show that even a primitive computer that is capable of implementing only SWAP gates, can be lifted to universal quantum computing, if aided with an appropriate quantum control of logarithmic size. Our results thus exemplify the information processing power brought forth by the quantum control system.
\end{abstract}

\date{\today}
\maketitle

\section*{Introduction}
Universal sets of gates for quantum computation are discrete sets of elementary unitary transformations which can be used to approximate any other unitary operation. Starting with Deutsch’s seminal proof of the universality of three-qubit gates \cite{Deutsch}, the search for universal sets of gates has been established as a mature theme in the field of quantum computing, and a plethora of various other universal sets have been thereafter constructed \cite{gates1,gates2,gates3,gates4,gates5}. Universal sets can be analyzed from two complementary perspectives: a practical one, which focuses on the efficiency and feasibility of implementing the gates, with the aim of providing a firm basis for future quantum computers; and a more conceptual one, which investigates how various universal sets shed light on the qualitative differences between classical and quantum computation. The latter perspective is exemplified in the works of Shi and Aharonov \cite{shi,aharonov}, which show that the Toffoli gate, when supplemented with the Hadamard gate, constitutes a universal set: as the Toffoli gate is universal for classical computation, one can see this result as highlighting the importance of the Hadamard gate (or, more generally, of the quantum Fourier transform) for the computational advantage brought forth by quantum computation.

Generally, if one wants to implement a unitary transformation on a target system, one can either act directly on that system (e.g. with a unitary device), or one can implement it indirectly by performing transformations and measurements on an ancillary control system. The latter method has been extensively studied in the context of quantum steering \cite{steer}, measurement based quantum computing \cite{mbqc1, mbqc2, mbqc3}, fusion-based quantum computing \cite{gimeno2014three,bartolucci2021fusion} and ancilla driven quantum computation (ADQC) \cite{ancilla1,kashefi2009twisted,proctor2017ancilla,halil2014minimum}. Motivated by the aforementioned results obtained by Shi and Aharonov, we may now ask the following question: \textit{can one achieve universal quantum computation by supplementing classical operations with a quantum control, and with a set of quantum gates acting solely on the latter?} More precisely, our aim is to consider computational models composed of classical sets of gates acting on a target system and an ancillary control system on which unitaries and measurements can be applied, and inspect whether they are sufficient for universal quantum computing. 

In this work we answer affirmatively to our question by constructing a computational model that consists of (i) a set of classical gates acting on $2n$ target bits, (ii) log($2n+1$) ancillary qubits which can coherently control the classical gates, and (iii) a set of quantum gates available on the control system. We show that this model can be used to perform universal quantum computation on $n$ qubits which are encoded in a subspace of the Hilbert space built upon the $2n$ target bits. The computation is executed via unitary transformations and measurements applied on the control system in a repeat-until-success manner \cite{lim2005repeat,lim2006repeat,ancilla2}. Interestingly, the set of classical gates that we start with consists only of local NOT and CNOT gates, which can be implemented by the so called \textit{parity computer}, thereby exhibiting a parallel between our work and the study of the ``computational power of correlations'' \cite{anders2009computational}. Namely, in the latter work the authors proved that a parity computer can be used to perform universal \textit{classical} computation, if supplemented with three-qubit GHZ states. On the other hand, in our work we show that a parity computer, when supplemented with an appropriate quantum control, can perform universal \textit{quantum} computation. As an addition, we show that the classical operations that we use can be alternatively implemented as local SWAP-gates, which implies that even a computer that can implement only such ``trivial class'' of reversible classical operations \cite{aaronson2015classification}, can be raised to universal quantum computation.

Returning back to the parallel with Shi's and Aharonov's results, our work shows an alternative method of turning a set of classical gates into a universal set for quantum computation: one simply needs to have access to an ancillary control degree of freedom on which certain unitary transformations (not necessarily the whole unitary group) can be implemented. Consequently, our results can be seen as a contribution to recent developments that show the information-theoretic advantage brought forth by the possibility of coherently controlling (quantum) operations, such as the coherent control of orders \cite{araujo2014computational}, directions \cite{guerin2016exponential}, communication channels \cite{ebler2018enhanced,guerin2019communication,abbott2020communication,rubino2021experimental}, and tasks and addressing in quantum networks \cite{miguel2020genuine}.

\section*{Description of the model}
The aim of this work is to construct a model of quantum computation via quantum controlled classical operations. 
The model will consist of a ``control system'' and of a ``target system'', where the former controls which operations are to be applied on the latter. 
In this section we will describe our model and introduce the main concepts, whereas the proof of universality will be given in the subsequent section.

\textit{Classical operations --} In this subsection we will state the precise meaning of ``classical operations''. Let us assume that the target system is a $2^n$-dimensional quantum system, i.e. a collection of $n$ qubits with its Hilbert space spanned by states $\left\{\ket{\vec{x}}\equiv \ket{x_1...x_n}, \forall x_i=0,1\right\}$; the latter basis will be referred to as the \textit{classical basis}. 
Moreover, let us assume that instead of having available the full unitary group $U(2^n)$ acting on the $n$ qubits, we have at disposal only a finite subset $\mathcal{G} \subset U(2^n)$. 
We will say that the operations $G_i \in \mathcal{G}$ are \textit{classical} if they are all equivalent to permutation matrices in the classical basis, i.e. if $\mathcal{G} \subseteq S_{2^n}$, where $S_{2^n}$ is the symmetric group on $2^n$ elements (we are thus considering only \textit{reversible} classical transformations). 
For example, for a single qubit ($n=1$), the only existing classical operations are the identity $\mathbb{1}$ and the NOT-gate, which act as $\mathbb{1}\ket{x}=\ket{x}$ and NOT$\ket{x}=\ket{x\oplus 1}$.
The reason we call the transformations pertaining to $\mathcal{G}$ classical is that by restricting the set of available transformations to $\mathcal{G}$, our target system behaves effectively as a classical system, i.e. the $n$ qubits behave effectively as $n$ bits\footnote{In other words, the limitation on the available transformations superselects quantum theory on $n$ qubits into an effective classical theory on $n$ bits.}. 
More precisely, each state $\ket{\vec{x}}$ of the $n$ qubits can be regarded as a state $\vec{x}$ of $n$ bits, and each operator $G_i \in \mathcal{G}$ acting on the qubits corresponds to a reversible function $\vec{f}^{(i)}\left(\vec{x}\right)$ of the $n$ bits, where $\ket{{\vec{f}^{(i)}\left(\vec{x}\right)}}=G_i\ket{\vec{x}}$. 
Additionally, all projective measurements in the classical basis correspond to ``classical measurements'' of the bits. 
Therefore, our system, together with the set of classical operations $\mathcal{G}\subseteq S_{2^n}$, can be used only to perform classical computation (if the set $\mathcal{G}$ is rich enough, one may be able to do universal classical computation on $n$ bits or on a subset thereof). Due to the above, we will throughout the manuscript interchangeably use the words ``bits'' and ``qubits'' when referring to the target system.
One may generalize the so far introduced definitions to mixed states and to generic completely positive and trace preserving (CPTP) maps, where classical operations would correspond to stochastic matrices (and would thus also include irreversible transformations); however, in this work we will focus only on pure states and on unitary transformations/permutations, in order to avoid unnecessary complications.

\textit{Control system --} 
We here introduce the control system as a $|\mathcal{G}|$-dimensional quantum system with its Hilbert space spanned by states $\left\{\ket{G_i},\forall G_i\in \mathcal{G} \right\}$; the latter basis of the control system will be referred to as the computational basis.
The role of the control system is to control the operations $\mathcal{G}$ that act on the target system, i.e. the interaction between the control and target systems is represented as:
\begin{equation}
\ket{G_i}\otimes \ket{\vec{x}} \rightarrow  \ket{G_i}\otimes G_i\ket{\vec{x}}.
\end{equation}

As opposed to the target system, we will not impose any restriction on the availability of operations acting on the control system, and will thus assume that any unitary operator can be implemented on it.

\begin{figure}[t]
\includegraphics[width=\linewidth]{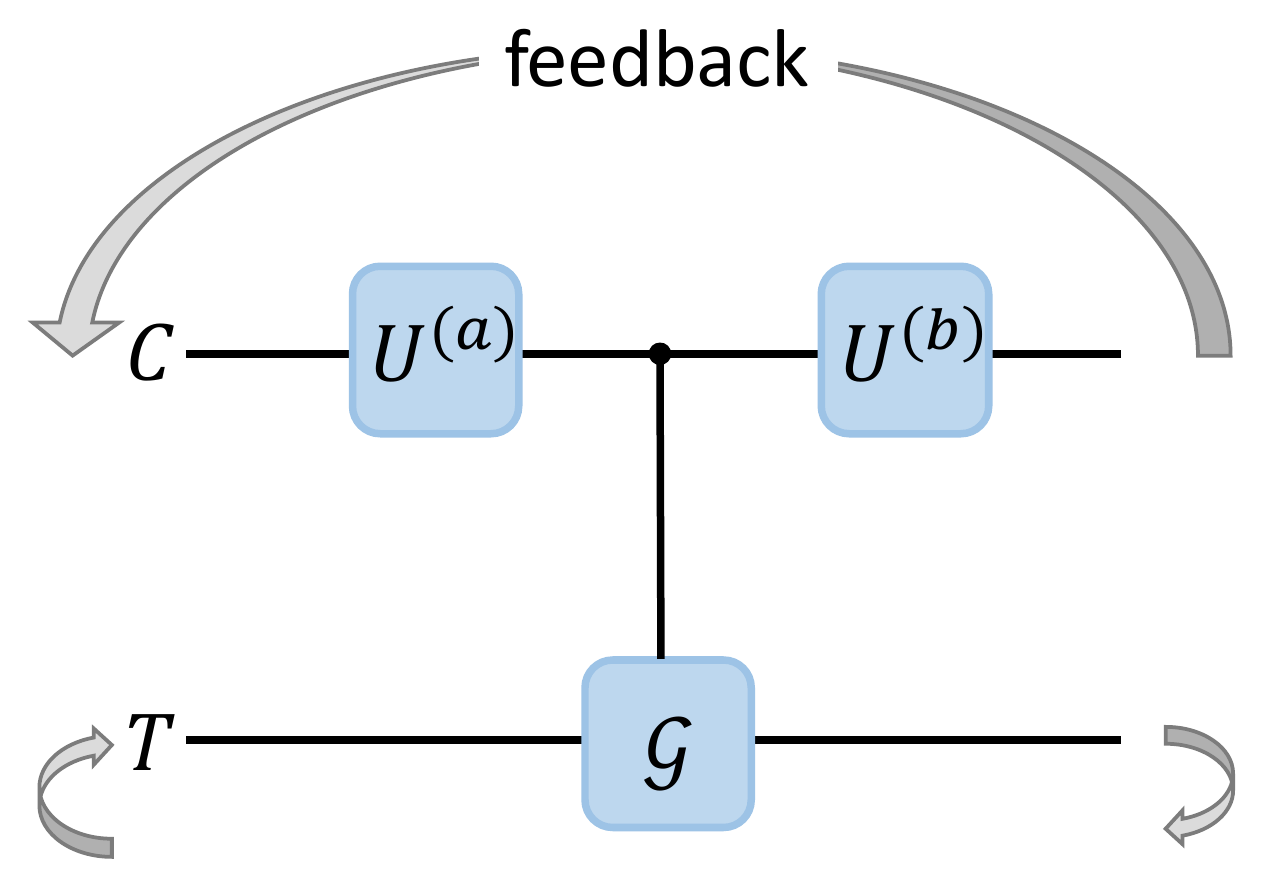}
\caption{The figure represents the general scheme for the implementation of transformations that lie outside of the classical set $\mathcal{G}$. The classical operations acting on the target system $T$ are coherently controlled by the control system $C$. $U^{(a)}$ and $U^{(b)}$ are generic unitaries acting on the control system. The arrows signify potential iterations of the procedure.}
\centering
\label{fig1}
\end{figure}

Now we briefly show a way in which the control system can be used to implement transformations on the target system, that lie outside of the pregiven classical set $\mathcal{G}$. Let us suppose that the initial (control+target) system is in state $\ket{G_0}\otimes \ket{\vec{x}}$, for arbitrary $G_0 \in \mathcal{G}$ and arbitrary $\vec{x}$. The general scheme, which is illustrated in Figure \ref{fig1} is partitioned in the following three steps.

\textbf{(a)} First, we start by acting with a unitary transformation $U^{(a)}$ on the control system, and we then let the control and target systems interact, leading to: 
\begin{equation}
\rightarrow \sum_i u^{(a)}_{i0} \ket{G_i}\otimes G_i\ket{\vec{x}},
\end{equation}
where $u^{(a)}_{i0}$$=\bra{G_i}U^{(a)}\ket{G_0}$.

\textbf{(b)} Next, we apply a unitary operator $U^{(b)}$ on the control system, resulting in:
\begin{equation}
\rightarrow \sum_j \ket{G_j}\otimes O^{(ab)}_j \ket{\vec{x}},   
\end{equation}
where $O^{(ab)}_j\equiv \sum_i  u^{(b)}_{ji} u^{(a)}_{i0} G_i$.

\textbf{(c)} Finally, we perform a projective measurement $\left\{\ket{G_j}\bra{G_j}, \forall j \right\}$ on the control system. For each outcome ``$j$'' the target system is thus postselected (up to normalization) into state $O^{(ab)}_j \ket{\vec{x}}$.

Therefore, the scheme (a)-(c) enables the probabilistic implementation of target system transformations $O^{(ab)}_j$, which may lie outside of $\mathcal{G}$. By reiterating this procedure one can probabilistically implement products of transformations $O^{(ab)}_j$, thereby forming the semigroup $\mathcal{L}\left[\mathcal{G}\right]$, which is defined as:
\begin{equation}
\mathcal{L}\left[\mathcal{G}\right]\equiv <O^{(ab)}_j, \forall U^{(a,b)}\in U(|\mathcal{G}|); j=1,...,|\mathcal{G}|>,   
\end{equation}
where $<S>$ indicates the semigroup generated by set $S$, and $O^{(ab)}_j = \sum_i  u^{(b)}_{ji} u^{(a)}_{i0} G_i$.
We will say that the set $\mathcal{G}$ \textit{can be lifted} to a (semi)group $\tilde{\mathcal{G}}$ with the aid of a control system, if $\tilde{\mathcal{G}} \subseteq \mathcal{L}\left[\mathcal{G}\right]$, that is, every element $g \in \tilde{\mathcal{G}}$ can be implemented by iterating procedure (a)-(c), in a repeat-until-success manner \cite{lim2005repeat,lim2006repeat,ancilla2}. Notice that the notion of lifting a set $\mathcal{G}$ to the (semi)group $\tilde{\mathcal{G}}$ can in principle be defined for any subset $\mathcal{G} \subseteq U(2^n)$; in this work we are however explicitly interested in $\mathcal{G}$ being a set of classical transformations, i.e. $\mathcal{G} \subseteq S_{2^n}$.

Recall that our goal here is to show the possibility of achieving universal quantum computation via quantum controlled classical operations. Now we are finally ready to formalize the latter problem in terms of the following question: can we find a set of classical transformations $\mathcal{G} \subseteq S_{2^n}$ that can be lifted to the $k$-qubit unitary group, i.e. $U(2^k) \subseteq \mathcal{L}\left[\mathcal{G}\right]$, for some $k \in \mathbb{N}$? Moreover, how large does $k$ need to be, i.e. is it possible to achieve linear scaling $k=\mathcal{O}(n)$?
In the next paragraph we will provide some preliminary steps in the analysis of this problem.

\textit{Preliminary steps --}
Our goal is to find a classical set $\mathcal{G}$ which can be lifted to the unitary group $U(2^k)$ acting on $k$ qubits with the aid of a control system. Notice that in order to show that $U(2^k) \subseteq \mathcal{L}\left[\mathcal{G}\right]$, it is sufficient to show that $\mathcal{L}\left[\mathcal{G}\right]$ contains any universal set of gates. The universal set that we are going to consider is the \textit{standard} set of gates $\{H, T, \text{CNOT}\}$ \cite{standard}, where $H$ and $T$ are the Hadamard and phase gate, which in the computational basis read as
\begin{equation}
H =
\frac{1}{\sqrt{2}} \begin{pmatrix} 
1 & 1 \\
1 & -1
\end{pmatrix},
\quad T =
\begin{pmatrix} 
1 & 0 \\
0 & e^{i\pi/4}
\end{pmatrix},
\end{equation}
and CNOT is a two-qubit gate which acts as $\text{CNOT} \ket{x_1x_2}=\ket{x_1 (x_2 \oplus x_1)}$.

We will start with the attempt of constructing the single-qubit gates $H$ and $T$ within our model (which would enable the construction of any other $U(2)$ transformation). Let us tentatively take the target system to be a single qubit, with the set of available classical operations being the maximal one, i.e. $\mathcal{G}=\left\{\mathbb{1},\text{NOT}\right\}$. 
As there are only two available transformations, the control system is two-dimensional, and thus also a qubit. 
By iterating procedure (a)-(c) described in the previous paragraph, one can implement only transformations of the form $\sim \left(\beta_1 \mathbb{1} + \beta_2 \text{NOT}\right)$ on the target system, where $\beta_{1,2}$ are accordingly normalized coefficients. Notice that all of these transformations mutually commute, which immediately points to the impossibility of implementing the required gates $H$ and $T$, as the latter do not commute. Therefore, if the target system is a single qubit, the set of classical transformations cannot be lifted to the full group $U(2)$. Nevertheless, it may be possible to use a target system that consists of more qubits, say $k$ of them, with the available classical set being $\mathcal{G} \subseteq S_{2^k}$, and to construct the single-qubit group $U(2)$ in a two-dimensional subspace of the $k$ qubits' Hilbert space. Formally, in order to see this possibility, note that the (maximal) group of classical operations $S_{d}$ is represented on the target's Hilbert space as the group of $d \times d$ permutation matrices, thus it can be decomposed into the trivial 1-dimensional irreducible representation and its complementary $(d-1)$-dimensional irreducible representation, the latter being known as the ``standard representation'' \cite{fulton2013representation}. The Burnside theorem then implies that the standard representation spans the set of all $(d-1)$-dimensional matrices \cite{burn}. Consequently, if the target system consists of $k$ qubits, together with the set of available transformations being $\mathcal{G}=S_{2^k}$, we expect the possibility of implementing the unitary group $U(2^k-1)$ on the corresponding irreducible subspace of the $k$-qubits' space. This implies that if we consider the target system to be composed of two qubits, together with the set of available operations being $\mathcal{G}=S_3 \subseteq S_4$, we expect the possibility of building any unitary transformation $U(2)$ on a two-dimensional subspace of the two-qubit Hilbert space. 
The subspace in which the constructed $U(2)$ acts will be called the \textit{logical subspace}, whereas the system that corresponds to this subspace will be called the \textit{logical qubit}. 
In what follows, we will fix the classical set $\mathcal{G}$ and specify the logical subspace.

\textit{Fixing the set $\mathcal{G}$ and encoding the logical qubits --}
As argued in the previous paragraph, we need to take at least two target qubits and encode the logical qubit into a two-dimensional subspace. We will choose the set of classical operations to be the following set of single-qubit and two-qubit classical gates:
\begin{equation}
\mathcal{G}=\left\{\mathbb{1}, \text{NOT}_1, \text{CNOT} \right\},
\end{equation}
where $\text{NOT}_1$ implements the NOT-gate on the first qubit, i.e. $\text{NOT}_1 \ket{x_1x_2}=\ket{(x_1 \oplus 1) x_2}$, and the CNOT-gate acts as $\text{CNOT} \ket{x_1x_2}=\ket{x_1 (x_2 \oplus x_1)}$. Notice that the NOT and CNOT gates do not commute, which hints to the possibility of implementing the group $U(2)$ in a certain subspace. 
Throughout the letter, we will for simplicity use the notation  $G_1=\text{NOT}_1$, and $G_2=\text{CNOT}$.

Now we will specify the logical subspace, i.e. the subspace in which the logical qubit will be encoded. We choose the computational basis $\left\{\ket{0}_L, \ket{1}_L\right\}$ of the logical qubit to be encoded as:
\begin{equation}
\begin{split}
&\ket{0}_L\equiv \frac{1}{\sqrt{2}}\left(\ket{10}-\ket{11} \right),\\
&\ket{1}_L\equiv \frac{1}{\sqrt{2}}\left(\ket{00}-\ket{01}  \right).
\end{split}
\end{equation}

By short inspection one can see that the classical transformations $G_1$ and $G_2$ act on the computational states of the logical qubit as follows:
\begin{equation}
\begin{split}
&G_1 \ket{y}_L=\ket{y \oplus 1}_L,\\
&G_2 \ket{y}_L=-(-1)^y \ket{y}_L,
\end{split}
\end{equation}
for $y=0,1$. 
Therefore, the transformations act on the logical qubit as $G_1\cong X$, and $G_2\cong -Z$, where $X$ and $Z$ are Pauli operators. Since the Pauli-$Y$ operator can be straightforwardly obtained as $G_2G_1\cong -iY$, we see that due to the particular way of encoding the logical qubit, the classical operations $\mathcal{G}$ furnish automatically the whole Pauli group, up to global phases. 
More precisely, the Pauli group for a single qubit is $P=\left\{e^{i\theta \frac{\pi}{2}}\sigma_j |\quad \theta,j=0,1,2,3 \right\}$, where $\sigma=(\mathbb{1},X,Y,Z)$. 
Next, consider the quotient of $P$ with its subgroup $Z_4 \cong \left\{e^{i\theta \frac{\pi}{2}}\mathbb{1} |\quad \theta=0,1,2,3\right\}$, which reads $P/Z_4=\left\{\left[\sigma_j\right] |\quad j=0,1,2,3 \right\}$, where $\left[\sigma_j\right]=\left\{e^{i\theta \frac{\pi}{2}}\sigma_j, \theta=0,1,2,3\right\}$. 
From the above considerations it follows that the set of classical operations $\mathcal{G}$ acting on 2 qubits furnishes automatically the group $P/Z_4$ acting on the logical qubit.

We will analogously construct $n$ logical qubits from $2n$ target ones, and assume the availability of the transformations $\mathcal{G}$ on each pair. Let us label with $G^{(l)}_i$ the transformation that acts as $G_i$ on the $l$-th target pair $(2l-1,2l)$, and trivially on the rest of the qubits, i.e. $G^{(l)}_i \equiv \mathbb{1}^{\otimes (2l-2)}\otimes G_i \otimes \mathbb{1}^{\otimes (2n-2l)}$. The total set of available transformations is then:
\begin{equation}\label{classical set G}
\mathcal{G}= \bigcup_{l=1}^{n} \left\{ G^{(l)}_1,G^{(l)}_2\right\} \cup \left\{\mathbb{1}\right\}.
\end{equation}

Since the latter set contains $(2n+1)$ elements, the control system has to be $(2n+1)$-dimensional, or in other words, a collection of log($2n+1$) qubits. As we will see in the next section, in order to lift the set $\mathcal{G}$ to the unitary group $U(2^n)$, we will not need to assume the availability of the whole group $U(2n+1)$ acting on the control system, but only of a subgroup of the latter. 

\begin{figure}[t]
\includegraphics[width=\linewidth]{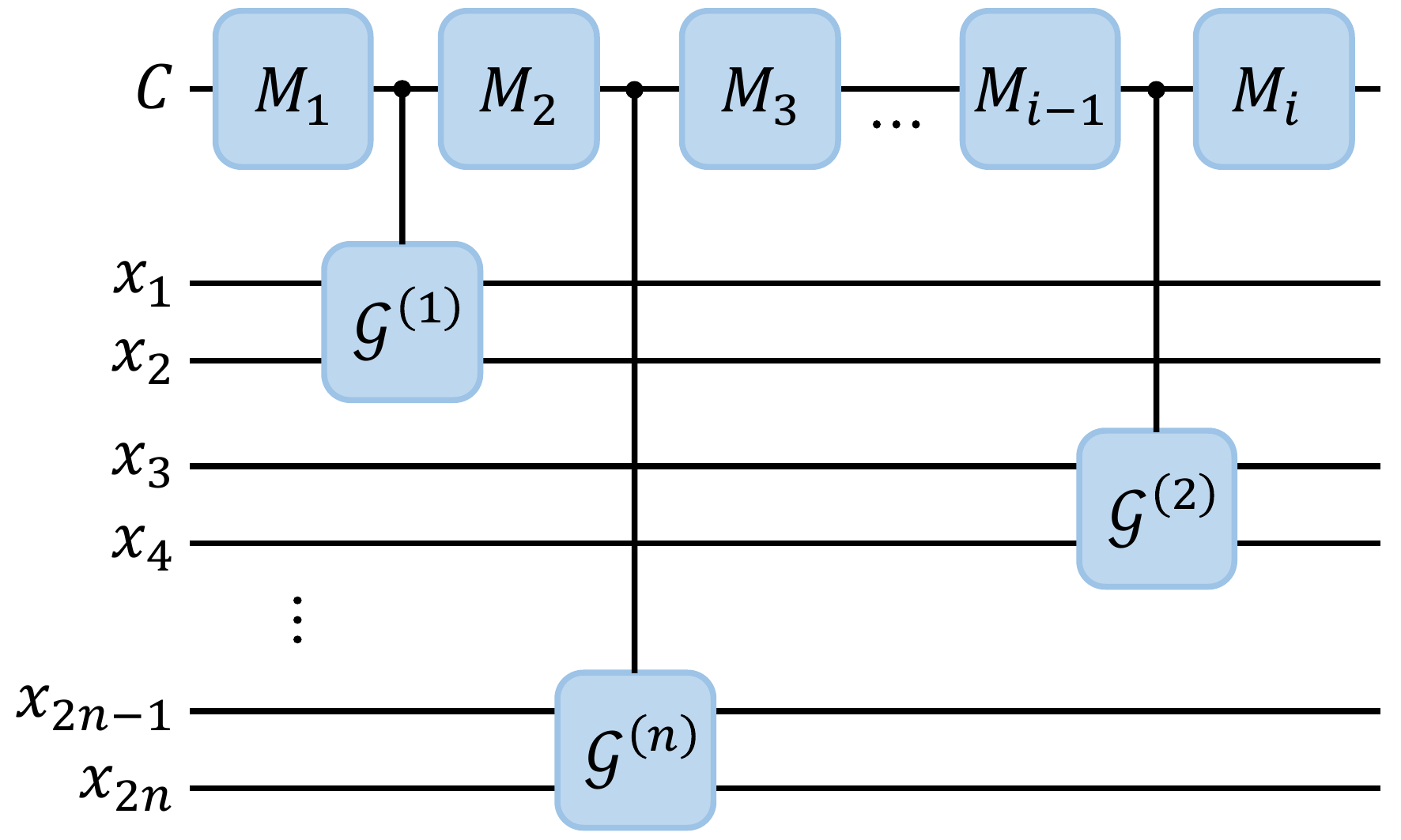}
\caption{Our model of computation which consists of $2n$ target qubits and a $(2n+1)$-dimensional control system $C$. The classical operations $\mathcal{G}$ are local on each qubit pair, and are coherently controlled by $C$. The operations $M_{i}$ represent unitary transformations and/or projective measurements applied on the control system.}
\centering
\label{fig2}
\end{figure}

To recapitulate, our computational model, which is pictured in Figure \ref{fig2}, consists of: (i) the target system that consists of $n$ qubit pairs, (ii) the set of classical transformations $\mathcal{G}$ acting on the target system, (iii) a $(2n+1)$-dimensional control system, and (iv) a subgroup of $U(2n+1)$ available on the control system. In the next section we will show that this model can be used to perform universal quantum computation. 

\section*{Universal quantum computation}\label{uqc}
In this section we will show that the model introduced in the previous section admits the implementation of the standard set of gates and thus enables universal quantum computation. Each gate will be implemented by iterating the scheme (a)-(c) presented in the previous section, and illustrated in Figure \ref{fig1}. However, before proceeding with the implementation of the elementary gates, we need to show how to initialize the $2n$ target qubits into a logical qubits' state, say $\ket{0}_L^{\otimes n}$, that is henceforth ready for computation.

\textit{Initialization --} Let us assume that each pair of qubits is prepared in the classical state $\ket{10}$, i.e. that the joint state of the target is initially $\ket{10}^{\otimes n}$. We will now take one of the pairs (for simplicity, we will omit the pair index) and show how to initialize it to state $\ket{0}_L$ (the procedure for the rest of the qubits then follows analogously).

Let us take the control system to be initially prepared in state $\ket{\mathbb{1}}$. 
The initialization is partitioned in three steps as follows:

(a) We apply a Hadamard gate on the control system in the $\left\{\ket{\mathbb{1}},\ket{G_2}\right\}$ subspace in order to prepare it in state $\frac{1}{\sqrt{2}}(\ket{\mathbb{1}} + \ket{G_2})$. We then let the control and target interact, which leads to:
\begin{equation}\label{in1}
\begin{split}
\rightarrow &\frac{1}{\sqrt{2}}\left(\ket{\mathbb{1}}\otimes \mathbb{1} \ket{10}+\ket{G_2}\otimes G_2 \ket{10}\right)\\
&=\frac{1}{\sqrt{2}}\left(\ket{\mathbb{1}}\otimes \ket{10}+\ket{G_2}\otimes\ket{11}\right).
\end{split}
\end{equation}

(b) Once again, we apply a Hadamard gate (in the same subspace as in the previous step) on the control system, resulting in:
\begin{equation}\label{in2}
\rightarrow \frac{1}{2}\left[\ket{\mathbb{1}}\otimes \left(\ket{10}+\ket{11}\right)+\ket{G_2}\otimes \left( \ket{10}-\ket{11}\right)\right].
\end{equation}

(c) Finally, we perform a projective measurement on the control system in the computational basis:
\begin{itemize}
    \item if the outcome is ``$G_2$'', then the postselected state of the target system is $\frac{1}{\sqrt{2}}\left(\ket{10}-\ket{11}\right)=\ket{0}_L$ and the initialization is complete;
    \item if the outcome is ``$\mathbb{1}$'', the state of the target system is $\frac{1}{\sqrt{2}}\left(\ket{10}+\ket{11}\right)$. 
    We then perform a ``classical measurement'' on the target system, i.e. a projective measurement with projectors $\left\{\ket{x_1x_2}\bra{x_1x_2}, \forall x_1,x_2=0,1 \right\}$, which sets the target system either in state $\ket{10}$ or in state $\ket{11}$. 
    Regardless of the outcome of the latter measurement, we reiterate the whole procedure (a)-(c).
\end{itemize}

We thus keep reiterating points (a)-(c) until the measurement in step (c) outputs ``$G_2$''; it is straightforward to see that the target system is then set in state $\ket{0}_L$ or $-\ket{0}_L$ and the initialization is thus complete, as the global phase is irrelevant.
The probability of achieving the desired outcome in $m$ runs grows exponentially fast towards certainty and is equal to $P(m)=\sum_{i=1}^{m}\left(\frac{1}{2}\right)^i=1-\left(\frac{1}{2}\right)^m$. 
We can analogously initialize $n$ qubit pairs which are prepared in the classical state $\ket{10}^{\otimes n}$, into the logical state $\ket{0}^{\otimes n}$; the probability of successfully doing the latter within $m$ iterations per qubit is $\left(1-\left(\frac{1}{2}\right)^m\right)^{n}$.

\textit{Single-qubit gates --}
We will now start by showing how to implement the single-qubit gates $H$ and $T$. 
We will thus again take one of the qubit pairs and omit the pair index for simplicity.

\textit{H gate.} We are going to show how to implement the Hadamard gate $H$ on a single logical qubit. 
Let us assume that the initial state of the joint (control+target) system is $\ket{G_1}\otimes \ket{\psi}_L$, where $\ket{\psi}_L$ is an arbitrary pure state pertaining to the logical subspace of the target system. 
The procedure consists of the following steps:

(a) First, notice that $H=\frac{1}{\sqrt{2}}\left(X+Z\right)$. Since we know that the operations $G_1$ and $G_2$ are equivalent to the Pauli operators $X$ and $-Z$ in the logical subspace, this motivates us to apply a Hadamard gate on the control system in the $\left\{\ket{G_1},\ket{G_2}\right\}$ subspace. The subsequent interaction between the control and target then leads to: 
\begin{equation}
\rightarrow \frac{1}{\sqrt{2}}\left(\ket{G_1}\otimes X\ket{\psi}_L-\ket{G_2}\otimes Z\ket{\psi}_L\right).
\end{equation}

(b) We proceed by again applying the Hadamard gate (in the same subspace as in step (a)) on the control system:
\begin{equation}
\rightarrow \frac{1}{2}\left(\ket{G_1}\otimes \left(X-Z\right)\ket{\psi}_L+\ket{G_2}\otimes \left(X+Z\right)\ket{\psi}_L\right).
\end{equation}

(c) Finally, we measure the control system in the computational basis: if the outcome is ``$G_1$'', the target system is in state $\frac{1}{\sqrt{2}}(X-Z)\ket{\psi}_L=\tilde{H} \ket{\psi}_L$, where we defined the unitary transformation $\tilde{H} \equiv \frac{1}{\sqrt{2}}(X-Z)$; on the other hand, if the outcome is ``$G_2$'', the target system is in state $\frac{1}{\sqrt{2}}(X+Z)\ket{\psi}_L=H\ket{\psi}_L$ and the procedure is terminated. 
Notice that the transformations $H$ and $\tilde{H}$ can be obtained from the Pauli operators $X$ and $Z$ via a $\pi/2$-rotation around the y-axis, which implies that they are elements of the Pauli group (up to global phases), i.e. $G_H \equiv \left\{[\mathbb{1}], [H], [\tilde{H}], [Y]\right\}\cong P/Z_4$. 
This means that for the ``wrong'' outcome ``$G_1$'' we can reiterate the same procedure (a)-(c) and obtain the target system in state $\tilde{H}\tilde{H}\ket{\psi}_L$ or $H\tilde{H}\ket{\psi}_L$, where $\tilde{H}\tilde{H}=\mathbb{1} \in G_H$ and $H\tilde{H}=iY \in G_H$. We can continue reiterating the procedure and after an exponentially short number of trials the desired gate $H$ is obtained. In the Supplementary Information we show that the probability of implementing the required gate $H$ within $m$ iterations is $P(m)=1-\left(\frac{1}{2}\right)^{\lceil m/2 \rceil}$.

\textit{T-gate.} The implementation of the $T$-gate is analogous to the implementation of the Hadamard gate and proceeds as follows.
Let us assume that the initial state of the whole system is $\ket{\mathbb{1}}\otimes \ket{\psi}_L$, where $\ket{\psi}_L$ is again an arbitrary logical qubit state. As before, the procedure is partitioned in three steps.

(a) Note that $T=e^{i\pi/8}\left(\cos(\pi/8)\mathbb{1}-i\sin(\pi/8)Z\right)$. 
This motivates us to act on the control system with the unitary transformation $HTH$, where $H$ and $T$ are the Hadamard gate and $T$-gate in the $\left\{\ket{\mathbb{1}},\ket{G_2}\right\}$ subspace, thereby setting the control in state $e^{i\pi/8}\left(\cos(\pi/8)\ket{\mathbb{1}} - i\sin(\pi/8)\ket{G_2}\right)$. The subsequent control-target interaction then leads to the following state (up to a global phase): 
\begin{equation}
\rightarrow \cos(\pi/8)\ket{\mathbb{1}}\otimes \mathbb{1}\ket{\psi}_L -i\sin(\pi/8)\ket{G_2}\otimes G_2\ket{\psi}_L.   
\end{equation}

(b) Next, we apply a Hadamard gate on the control system (in the same subspace as in the previous step), thereby obtaining:
\begin{equation}
\begin{split}
&(\ket{\mathbb{1}}\otimes \left(\cos(\pi/8) \mathbb{1}-i\sin(\pi/8) G_2\right)\ket{\psi}_L+\\
&+\ket{G_2}\otimes \left(\cos(\pi/8) \mathbb{1}+i\sin(\pi/8) G_2 \right)\ket{\psi}_L).
\end{split}
\end{equation}

(c) Finally, we perform a measurement on the control system in the computational basis: if the outcome is ``$\mathbb{1}$'', the target system is set (up to a global phase) in state $T^{\dagger}\ket{\psi}_L$, whereas if the outcome is ``$G_2$'', the state of the target system is (up to a phase) $T\ket{\psi}_L$. 
Analogously to the implementation of the Hadamard gate, we can keep reiterating the procedure (a)-(c) until we get the desired gate. The group of transformations that is generated by the iterative procedure is equivalent (up to global phases) to the cyclic group $Z_8$, i.e. $G_T\equiv \left\{T^j,j=0,...,7\right\} \cong Z_8$. Notice also that $T=ZT^5$; thus, if at some step during the iteration of the procedure we obtain the unitary $T^5$, then we can apply the $Z$-gate (i.e. the classical gate $G_2$), and obtain the desired operation. Taking the latter into account, in the Supplementary Information we prove that the probability of generating the required gate $T$ within $m$ iterations is $P=1-\left(\frac{1}{2}\right)^{\lceil m/2 \rceil}$.

The possibility of implementing the $H$ and $T$ gates implies the possibility of implementing any single-qubit gate; in order to perform an arbitrary unitary operation on $n$ logical qubits, we still need to show how to implement the two-qubit CNOT-gate.

\textit{CNOT-gate.}
Let us assume that we want to implement the CNOT gate on the $k$-th and $l$-th logical qubits, where the latter are in some arbitrary joint state $\ket{\phi}_L$. As we are now dealing with two pairs of target qubits, we will use explicitly the pair indices $k$ and $l$ to indicate the state of the control system: for example, if the control system is prepared in state $\ket{G^{(k)}_i}$, then the control and target interact as:
\begin{equation}
\ket{G^{(k)}_i} \otimes \ket{\phi}_L \rightarrow \ket{G^{(k)}_i}\otimes (G_i \otimes \mathbb{1})\ket{\phi}_L,
\end{equation}
whereas, if the control system is in state $\ket{G^{(l)}_i}$, the interaction leads to:
\begin{equation}
\ket{G^{(l)}_i} \otimes \ket{\phi}_L \rightarrow \ket{G^{(l)}_i}\otimes (\mathbb{1} \otimes G_i)\ket{\phi}_L.
\end{equation}

\begin{figure}[t]
\includegraphics[width=\linewidth]{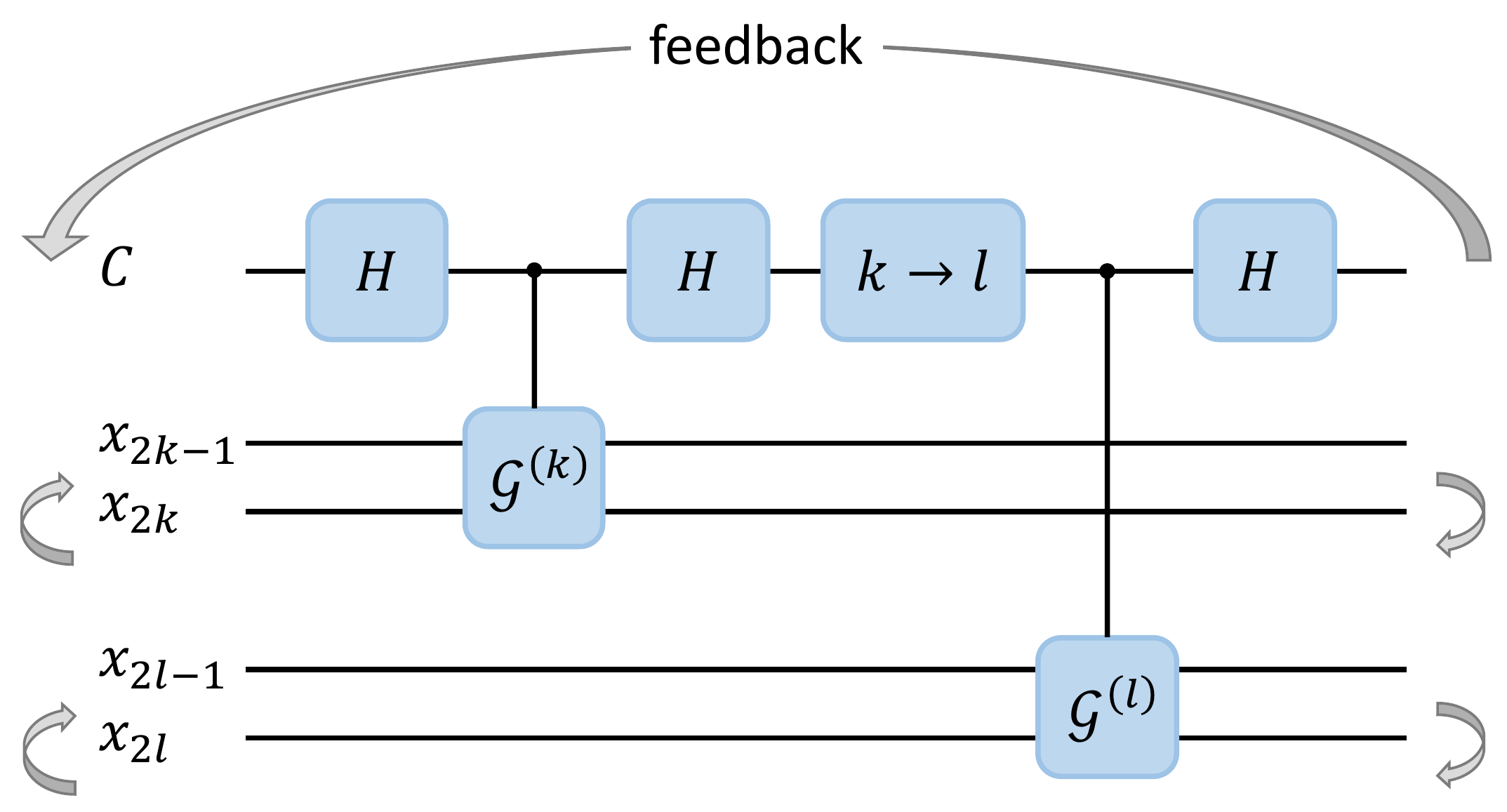}
\caption{Implementation of the CZ gate acting on the $k$-th and $l$-th logical qubits. The gate labelled with ``$k \rightarrow l$'' stands for the unitary transformation that acts as $\ket{G^{(k)}_i} \rightarrow \ket{G^{(l)}_i}$, whereas $H$ stands for the Hadamard gate. The arrows indicates that the procedure may be reiterated conditioned on the measurement outcomes.}
\centering
\label{cnot}
\end{figure}

Next, note that the CNOT gate can be implemented using the CZ gate as CNOT$=(\mathbb{1}\otimes H)\text{CZ}(\mathbb{1}\otimes H)$, where CZ$=\ket{0}\bra{0} \otimes \mathbb{1}+\ket{1}\bra{1} \otimes Z$. 
Since we know how to implement the $H$-gate, it is sufficient to prove the possibility of implementing the CZ-gate. 
Moreover, the CZ-gate can be decomposed as CZ$=\frac{1}{2}\left(\mathbb{1}\otimes\mathbb{1}+\mathbb{1}\otimes Z+Z \otimes\mathbb{1}-Z \otimes Z\right)$. 
In order to implement it within our model, we will use the following scheme, which is pictured in Figure \ref{cnot}.

(a) Let us assume that the joint (control+target) system is initially in state $\ket{G^{(k)}_2}\otimes \ket{\phi}_L$. 
We apply a Hadamard gate on the control system in the $\left\{\ket{\mathbb{1}},\ket{G^{(k)}_2}\right\}$ subspace, and we let the control and target systems interact:
\begin{equation}
\rightarrow \frac{1}{\sqrt{2}}\left(\ket{\mathbb{1}} \otimes \left(\mathbb{1}\otimes\mathbb{1}\right)\ket{\phi}_L- \ket{G^{(k)}_2}\otimes \left(Z\otimes\mathbb{1}\right)\ket{\phi}_L\right).   
\end{equation}

(b) Once again, we apply the Hadamard gate on the control system (in the same subspace as before):
\begin{equation}
\begin{split}
&\frac{1}{2}(\ket{\mathbb{1}}\otimes \left(\mathbb{1}\otimes\mathbb{1}+Z\otimes\mathbb{1}\right)\ket{\phi}_L+\\
&\ket{G^{(k)}_2}\otimes \left(\mathbb{1}\otimes\mathbb{1}-Z\otimes\mathbb{1}\right)\ket{\phi}_L).
\end{split}
\end{equation}
Next, we apply a unitary transformation on the control system which acts as $\ket{G^{(k)}_i} \rightarrow \ket{G^{(l)}_i}$, in order for it to implement transformations on the $l$-th qubit in the subsequent interaction. 
We then let the control and target systems interact and apply a Hadamard gate in the $\left\{\ket{\mathbb{1}},\ket{G^{(l)}_2}\right\}$ subspace. 
The output state is thus:
\begin{equation}
\begin{split}
&\frac{1}{2\sqrt{2}}(\ket{\mathbb{1}}\otimes \left(\mathbb{1}\otimes \mathbb{1}+Z\otimes\mathbb{1}-\mathbb{1}\otimes Z+Z\otimes Z\right)\ket{\phi}_L+ \\
&\ket{G^{(l)}_2}\otimes \left(\mathbb{1}\otimes\mathbb{1}+Z\otimes\mathbb{1}+\mathbb{1}\otimes Z - Z \otimes Z\right)\ket{\phi}_L).
\end{split}
\end{equation}

(c) A final measurement of the control yields the target system either in state $\widetilde{\text{CZ}}\ket{\phi}$, or in state CZ$\ket{\phi}$, where we introduced the unitary operator $\widetilde{\text{CZ}}=\frac{1}{2}\left(\mathbb{1}\otimes\mathbb{1}+Z\otimes\mathbb{1}-\mathbb{1}\otimes Z+Z\otimes Z\right)$. 
By reiterating the procedure, the only new transformations which are generated are $\mathbb{1}$ and $\widetilde{\text{CZ}}'\equiv \text{CZ} \times \widetilde{\text{CZ}}$. 
It is easy to check that the resulting group $G_{CNOT} \equiv \left\{\mathbb{1}, \text{CZ}, \widetilde{\text{CZ}}, \widetilde{\text{CZ}}'\right\}$ is equivalent to $P/Z_4$. 
The latter implies that the probability of implementing the correct transformation after $m$ trials is the same as the one needed for implementing the $H$ gate, i.e. $P(m)=1-\left(\frac{1}{2}\right)^{\lceil m/2 \rceil}$, as proved in the Supplementary Information.

\textit{Measurements of the logical qubits --} We have shown that our model enables the implementation of any unitary transformation acting on $n$ logical qubits. At the end of the computation, one still needs to be able to read out the state of the qubits, i.e. to perform a projective measurement of the logical qubits in the computational basis $\left\{\ket{0}_L, \ket{1}_L\right\}$. Due to our simple definition of the logical subspace, the required measurement can be implemented straightforwardly as follows: we perform a (classical) projective measurement on the target qubits in the classical basis; measurement outcomes $\ket{10}$ and $\ket{11}$ correspond to the logical outcome ``0'', whereas outcomes $\ket{00}$ and $\ket{01}$ correspond to the logical outcome ``1'' (projective measurements on more logical qubits then follow analogously).

\textit{Computational cost --} Now we will briefly comment on the computational cost of our model. Consider that we want to simulate a quantum circuit of size $K$, i.e. a sequence of $K$ gates, each drawn from the standard set $\{H, T, \text{CNOT}\}$. As proved before, each elementary gate can be implemented within our model with probability $P(m)=1-\left(\frac{1}{2}\right)^{\lceil m/2 \rceil}$, where $m$ is the number of iterations of the general procedure (a)-(c). Furthermore, $n$ logical qubits can be initialized within $m_0$ iterations per qubit with probability $\left(1-\left(\frac{1}{2}\right)^{m_0}\right)^{n}$. Therefore, the total probability of correctly implementing the desired circuit is: 
\begin{equation}\label{comp cost}
P(m_0,...,m_K)=\left(1-\left(\frac{1}{2}\right)^{m_0}\right)^{n} \prod_{i=1}^{K} \left(1-\left(\frac{1}{2}\right)^{\lceil m_i/2 \rceil}\right),
\end{equation}
where $m_i$ is the number of iterations involved in the implementation of the $i$-th gate. Next, we want to estimate the number of iterations needed in order to achieve probability $1-\delta$, for some $\delta \ll 1$. In order to simplify the analysis, we will take the number of iterations spent on each gate to be equal, i.e. $m_i=m_j=m, \forall i,j$, and we will set this number to be $m=2m_0$. Inserting these specifications into equation \eqref{comp cost} and equating the latter to the desired probability of success we obtain the following condition:
\begin{equation}
\left(1-\left(\frac{1}{2}\right)^{m_0}\right)^{n+K}=1-\delta.  
\end{equation}

By expanding the latter equation up to first order in $\delta$, we get $m_0=\text{log}\left(\frac{K+n}{\delta}\right)$. Consequently, the total number $M$ of gates required to simulate a circuit of size $K$ is:
\begin{equation}
M = \alpha (2K+n)\text{log}\left(\frac{K+n}{\delta}\right),
\end{equation}
where the factor $\alpha = \mathcal{O}(1)$ takes into account the fact that each iteration involves around 3 to 8 gates applied to the control and target systems.

\section*{Parity and SWAP computing}
In the previous section we showed that one can efficiently perform universal quantum computation by coherently controlling (i.e. lifting) the classical set $\mathcal{G}$ introduced in equation \eqref{classical set G}. It is interesting to note that the operations pertaining to the available set $\mathcal{G}$ can be implemented by the so called \textit{parity computer}, the latter being a classical device that can implement only NOT and CNOT gates (and is capable of computing all parity-preserving affine transformations) \cite{anders2009computational, aaronson2015classification}. Our result thus implies that if one is given a parity computer operating on $2n$ bits, one can achieve full quantum computation by supplementing it with log($2n+1$) control qubits. This implication draws a parallel between our work and the study of the ``computational power of correlations'' \cite{anders2009computational}, as the latter shows that one can achieve universal \textit{classical} computation by aiding a parity computer with three-qubit GHZ-states.

We will now briefly explain how our result implies that even a computer that is only capable of implementing SWAP gates can be lifted to universal quantum computation\footnote{A SWAP gate is a two-qubit gate that acts as SWAP$\ket{x_1x_2}=\ket{x_2x_1}$.}. Namely, it is sufficient to note that the classical set $\mathcal{G}$ can be realized in an alternative manner: instead of taking the available operations to be single-qubit NOT-gates and two-qubit CNOT-gates, we can use SWAP gates, as follows. We will encode one logical qubit into a two-dimensional subspace of a Hilbert space of \textit{four} qubits, as:
\begin{equation}
\begin{split}
&\ket{0}_L\equiv \frac{1}{\sqrt{2}}\left(\ket{1000}-\ket{0100} \right),\\
&\ket{1}_L\equiv \frac{1}{\sqrt{2}}\left(\ket{0010}-\ket{0001}  \right).
\end{split}
\end{equation}

Let us define $G_1=\text{SWAP}_{13}\text{SWAP}_{24}$ and $G_2=\text{SWAP}_{12}$, where $\text{SWAP}_{ij}$ is a SWAP gate acting on the $i$-th and $j$-th qubits. It is straightforward to check that these gates act on the logical qubit as $G_1\cong X$ and $G_2\cong -Z$. Therefore, one can simply take all the results from the previous sections and prove the possibility of performing universal computation by coherently controlling SWAP gates (albeit, in this case, in order to perform computation on $n$ logical qubits, one requires $4n$ target qubits). Therefore, if one is given a device that can implement only SWAP gates on $4n$ bits (and is thus capable of generating the trivial class of reversible classical transformations according to the classification in \cite{aaronson2015classification}), one can lift it to universal quantum computation by aiding it with log($2n+1$) control qubits. We want to point out the sim-
ilarity between our conclusions and the ones provided in \cite{lau2016universal}, where the authors presented a deterministic scheme that uses SWAP gates for universal quantum computing with bosonic systems.

\section*{Conclusion and outlook}
Now we will summarize and discuss the results presented in this work. Our goal is to show that one can achieve universal quantum computation on a target system via the coherent control of classical operations. We started by restricting the set of available operations acting on the target system to be classical, and we showed how a control system can be used to implement transformations that lie outside of the available set. We proceeded by showing that a particular set of gates, which consists of local NOT and CNOT gates acting on $2n$ target qubits, can be lifted to the full unitary group acting on $n$ logical qubits. The Hilbert space of the control system is only $(2n+1)$-dimensional (i.e. composed of log$(2+1)$ qubits), and the transformations available on the latter involve only U(2) gates acting in various two-dimensional subspaces. Moreover, all elementary gates (including the initialization of the logical qubits) feature a repeat-until-success strategy, with probabilities of success that converge exponentially fast to 1 with the number of repetitions: more precisely, in order to simulate a circuit of size $K$ on $n$ qubits with probability $1-\delta$, one needs to act with $\mathcal{O}( (2K+n)\text{log}\left(\frac{K+n}{\delta}\right))$ gates on the control and target systems. Finally, after noting that the classical transformations $\mathcal{G}$ can be executed by a parity computer, we showed that our results imply that even a device that implements only SWAP gates on $4n$ bits can be raised to full quantum computing on $n$ qubits, by aiding it with log($2n+1$) control qubits.

The procedure of enlarging a subset $\mathcal{G}$ of available operations to a larger (semi)group $\tilde{\mathcal{G}}$ by using a control system was named \textit{lifting}, which may as well be defined for any subset $\mathcal{G}$ of the unitary group, and not only for classical transformations. It would thus be interesting to analyze the lifting procedure in more general terms: one may take an arbitrary subset of the unitary group and inspect the (semi)group that one obtains by using a control system. Furthermore, one should take into account the efficiency of the procedure, i.e. the dimension of the control system and the probability of the successful gates' implementations. Such an analysis would contribute to the understanding and unification of the multifaceted roles that the coherent control of operations plays in quantum information processing.

Additionally, it would be interesting to inspect potential experimental implementations of our computational scheme, e.g. in the context of light-assisted \cite{beige2000quantum, Duan2004ScalablePQ, reiserer2014quantum, lim2005repeat, hacker2016photon, 2004} and matter-assisted quantum computation \cite{bartlett2021deterministic}.

\section*{Acknowledgements}
We thank Hoi-Kwan (Kero) Lau for bringing our at-
tention to the related work presented in \cite{lau2016universal}. The authors thank Anton Zeilinger, Elizabeth Agudelo, Faraj Bakhshinezhad and Philip Taranto for valuable discussions. 
This work was supported by the Austrian Academy of Sciences (ÖAW). X.G. acknowledges support from the Austrian Academy of Sciences (ÖAW) and the Joint Centre for Extreme Photonics (JCEP). S.H. acknowledges support from the Austrian Science Fund (FWF) through BeyondC-F7112. B.D. acknowledges support from an ESQ Discovery Grant of the Austrian Academy of Sciences (ÖAW) and the Austrian Science Fund (FWF) through BeyondC-F7112.

\textbf{Note on author contributions.} S.H. and X.G. contributed equally to this work.

%\bibliography{refs}
%\bibliographystyle{unsrt}
%\bibliography{refs.bib}

%\section{References}
%\bibliographystyle{unsrt}
\bibliography{refs}

\clearpage
\newpage

%\appendix

\section*{Supplementary Information}

\subsection*{Probabilities of successfully implementing the elementary gates} \label{appA}
In this section we will compute the probability of successfully implementing each element of the standard set of gates using a finite number of iterations of the procedures described in the main text.

\textit{H and CNOT gates --} As described in the main text, the iteration of the procedures involved in the implementation of the $H$ and CNOT gates lead respectively to the generation of groups $G_H$ and $G_{CNOT}$, which are both equivalent to $P/Z_4$. 
Here we will focus on the group $G_H=\left\{[\mathbb{1}], [H], [\tilde{H}], [Y]\right\}$; the result for the CNOT-gate then follows analogously.

Our aim is to answer the following question: what is the total probability $P(m)$ of implementing the required gate $[H]$, if we have $m$ iterations of the procedure at disposal? 
Elementary probability theory implies that $P(m)=\sum_{i=1}^{m}P^{(i)}$, where $P^{(i)}$ is the probability of implementing the correct transformation after $i$ iterations, given that the transformation has not been implemented in the previous $(i-1)$ iterations. 
Let us suppose that the initial state of the target system is $\ket{\psi}$; the first iteration of the procedure described in the main text outputs the state $[H]\ket{\psi}$ or $[\tilde{H}]\ket{\psi}$, each with probability 1/2: thus $P^{(1)}=1/2$. 
In the case that the outcome of the latter iteration is $[\tilde{H}]$, we reiterate the procedure once more and obtain results $\ket{\psi}$ or $[Y]\ket{\psi}$, which implies that $P^{(2)}=0$. 
Another reiteration yields $[H]\ket{\psi}$ or $[\tilde{H}]\ket{\psi}$, each with probability 1/2: therefore $P^{(3)}=1/4$. 
It is then easy to generalize the latter and see that $P^{(i)}=0$ for even $i$, and $P^{(i)}=\left(\frac{1}{2}\right)^{\lceil i/2 \rceil}$ for odd $i$. 
The total probability is thus $P(m)=\sum_{j=1}^{\lceil m/2 \rceil}\left(\frac{1}{2}\right)^{j} = 1-\left(\frac{1}{2}\right)^{\lceil m/2 \rceil}$.

\begin{figure}[t]
\includegraphics[width=0.7\linewidth]{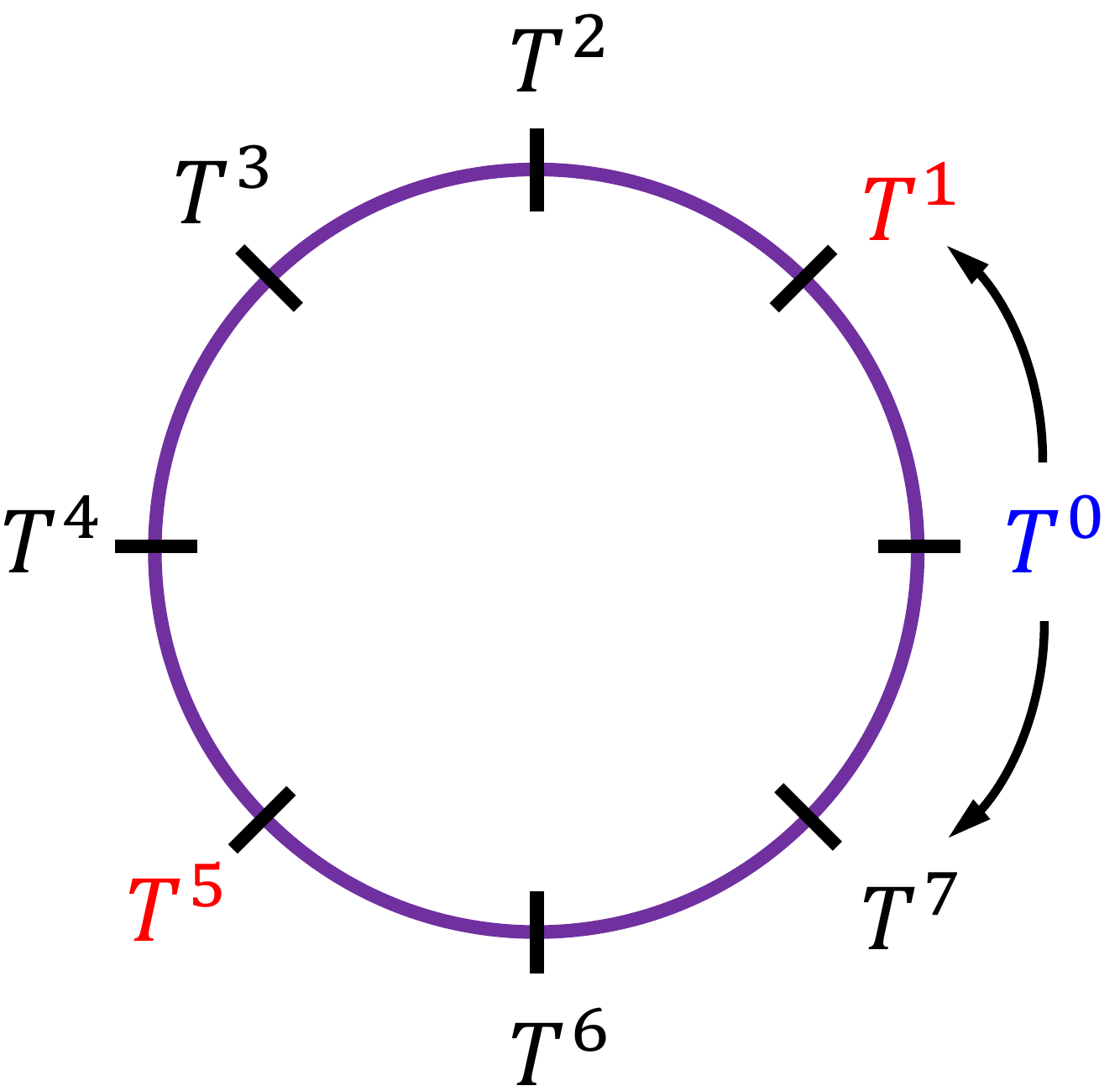}
\caption{Implementation of the $T$-gate illustrated as a random walk of a ``particle'' on eight uniformly distributed points on a circle. The walk starts at point $T^0$ and each iteration moves the particle by one step either in the clockwise or counterclockwise direction with uniform probability. 
The walk is terminated when the particle reaches points $T^1$ or $T^5$.}
\centering
\label{fig4}
\end{figure}

\textit{T-gate --} The procedure involved in the implementation of the T-gate generates the cyclic group $\left\{T^j,j=0,...,7\right\}$.
Every iteration of the procedure implements transformations $T$ or $T^{\dagger}$, each with probability 1/2. 
The overall process can thus be visualized as a random walk of a ``particle'' on 8 uniformly distributed points on a circle, as in Figure \ref{fig4}. The walk starts at point $T^0$ and each iteration generates one step in the clockwise or counterclockwise direction. 
Moreover, notice that $T=ZT^5$: this implies that if the particle finds itself at point $T^5$, we can easily obtain the desired gate $T$, as one of the available classical gates is isomorphic to the $Z$ gate. Our goal is thus to compute the total probability of the particle being at points $T^1$ or $T^5$ after $m$ steps of the walk. 
Similarly as in the previous paragraph, the total probability is $P(m)=\sum_{i=1}^{m}P^{(i)}$, where $P^{(i)}$ is the probability of the particle being in state $T^1$ or $T^5$ after $i$ steps, given that it has not passed through these states in the previous $(i-1)$ steps of the walk. 
Next, notice that in order for the particle to find itself in state $T^1$ or $T^5$ at the $i$-th step, it must necessarily have been either at $T^0$ or at $T^6$ at the $(i-1)$-th step. 
Therefore: 
\begin{equation}\label{appprob}
P^{(i)}=\frac{1}{2}\left(P^{(i-1)}_0+P^{(i-1)}_6\right), 
\end{equation}
where $P^{(i-1)}_j$ is the probability of the particle being located at point $T^j$ at the $(i-1)$-th step of the walk. 
Our goal now is to calculate $P^{(i-1)}_j$, for $j=0,6$.

Let us first define the space of normalized probabilistic states of the particle as a convex space St which can be embedded in an 8-dimensional vector space V, i.e. $\text{St} \subset$ V. Next, we introduce an orthonormal basis $\left\{\vec{e}_i, i=0,1,...,7 \right\}$, such that, if the particle is in the probabilistic state $\vec{p} \in \text{St}$, then $\vec{e}_k \cdot \vec{p}$ is the probability of it being located at point $T^k$. Moreover, the evolution of the particle is represented by a sequential application of a stochastic map $M: \text{St} \rightarrow \text{St}$.

Now, notice that we are interested exclusively in the walk of the particle through points $T^0$, $T^7$ and $T^6$; we will thus focus on the convex subspace $Q \subset \text{St}$, which is spanned by the basis vectors $\left\{\vec{e}_0, \vec{e}_7, \vec{e}_6\right\}$. With all of this said, it is easy to see that the evolution operator $M$, when restricted to the subspace $Q$, takes the following form (written in the basis $\left\{\vec{e}_0, \vec{e}_7, \vec{e}_6\right\}$):\\
$\quad \quad \quad \quad \quad \quad M_Q$ =
$\frac{1}{2} \begin{pmatrix} 
0 & 1 & 0\\
1 & 0 & 1\\
0 & 1 & 0
\end{pmatrix}$.\\

Note that the latter operator is not stochastic, as the walk can map the system outside of the (three-dimensional) domain that we are considering.

Now we are finally ready to compute $P^{(i-1)}_0$ and $P^{(i-1)}_6$. Namely, if the particle is initially in state $\vec{v}$, then the state after $m$ iterations of the walk is $(M_Q)^m\vec{v}$, \textit{given that} the particle has not been mapped outside of our three points in the previous steps. Therefore, since the particle is initially in state $\vec{e}_0$ (i.e. at point $T^0$), it follows that $P^{(i-1)}_j=\vec{e}_j \cdot (M_Q)^{i-1} \vec{e}_0$. Diagonalizing the matrix $M_Q$ and calculating its power yields:
\begin{equation}
\vec{e}_0 \cdot (M_Q)^{k} \vec{e}_0 = \vec{e}_6 \cdot (M_Q)^{k} \vec{e}_0 = \begin{cases} 
0, \quad &\text{for odd $k$};\\
\left(\frac{1}{2}\right)^{k/2+1}, \quad &\text{for even $k$.}
\end{cases}
\end{equation}

Together with equation \eqref{appprob}, the latter implies that $P^{(i)}=0$, for even $i$, and $P^{(i)}=\left(\frac{1}{2}\right)^{\lceil i/2 \rceil}$, for odd $i$.
Finally, we obtain $P(m)=\sum_{j=1}^{\lceil m/2 \rceil}\left(\frac{1}{2}\right)^{j} = 1-\left(\frac{1}{2}\right)^{\lceil m/2 \rceil}$.

\end{document}